\documentclass{iau}
\usepackage{graphicx}

\title[Multi-mode Cepheids and RR\thinspace Lyrae stars] 
{Multi-mode oscillations in classical Cepheids and RR\thinspace Lyrae-type stars}

\author[Pawe{\l} Moskalik] 
{Pawe{\l} Moskalik}

\affiliation{Copernicus Astronomical Centre, ul. Bartycka 18, 00-716 Warsaw, Poland
\\ email: {\tt pam@camk.edu.pl}}

\pubyear{2014}
\volume{301}  
\pagerange{1--8}
\jname{Precision Asteroseismology}
\editors{J.A. Guzik, W.J. Chaplin, G. Handler \& A. Pigulski, eds.}
\begin{document}

\maketitle

\begin{abstract}
I review different types of multi-mode pulsations observed in
classical Cepheids and in RR\thinspace Lyrae-type star. The
presentation concentrates on the newest results, with special
emphasis on recently detected nonradial oscillations.
\end{abstract}

\firstsection 

\section{Introduction}

Most of classical (Pop.\thinspace I) Cepheids and RR\thinspace
Lyrae-type stars are periodic, single-mode radial pulsators. The
long-term lightcurve modulation, observed in some RR\thinspace
Lyrae-type variables, was for many years the only known exception
from this simple picture. This phenomenon, known as the Blazhko
effect, was discovered a century ago (\cite[Blazhko 1907]{Blazhko},
\cite[Shapley 1916]{Shapley}) and still lacks a satisfactory
explanation. Current understanding of the Blazhko modulation has
been reviewed by \cite[Kov\'acs (2009)]{Kov09} and by Szab\'o (these proceedings) and will not be discussed here. Instead, I will
focus on various forms of multi-mode pulsations, i.e. pulsations
with several different oscillation modes simultaneously excited.


\section{Classical double-mode radial pulsators}


{\underline{\it F+1O double-mode Cepheids.}} The first double-mode
pulsators, U\thinspace TrA and TU\thinspace Cas, were identified
more than fifty years ago (\cite[Oosterhoff
1957a]{UtrA},\cite[b]{TUCas}). These two Cepheids pulsate
simultaneously in the two lowest radial modes -- the fundamental mode
(F) and the first overtone (1O). Double-mode Cepheids of this class
have period ratios of $P_1/P_0=0.694-0.746$. Almost 200 such stars
are currently known in the Galaxy and the Magellanic Clouds
(\cite[Soszy\'nski et al. 2008b]{LMCCep}, \cite[2010a]{SMCCep},
\cite[2011b]{BulgeCep}, \cite[2012]{SEP}, \cite[Marquette et al.
2009]{EROSCep}, \cite[Smolec \& Moskalik 2010]{SM10} and references
therein). A few have also been discovered in M33 (\cite[Beaulieu et
al. 2006]{M33Cep}) and most recently in M31 (\cite[Poleski
2013a]{M31Cep}).

{\underline{\it F+1O double-mode RR\thinspace Lyrae-type stars (RRd
stars).}} These variables are RR\thinspace Lyrae-type analogs of the
F+1O double-mode Cepheids. AQ\thinspace Leo, the prototype of the
class, was discovered by \cite[Jerzykiewicz \& Wenzel
(1977)]{AQleo}. Since then, nearly 2000 RR\thinspace Lyrae stars of
this type have been identified. They are observed in the Galactic
field (\cite[Wils 2010]{Wils}, \cite[Poleski 2013b]{GalRRd}), in
many Galactic globular clusters (e.g. \cite[Walker 1994]{M68RR},
\cite[Walker \& Nemec 1996]{IC4499RR}, \cite[Corwin et al.
2008]{M15RR}), in the Magellanic Clouds (\cite[Soszy\'nski et al.
2009]{LMCRR}, \cite[2010b]{SMCRR}, \cite[2012]{SEP}) and in nearby
dwarf spheroidal galaxies (e.g. \cite[Cseresnjes 2001]{SgrRR},
\cite[Clementini et al. 2006]{ForRR}, \cite[Bernard et al.
2009]{TucRR}). In most stellar populations, their period ratios are
in a very narrow range of $P_1/P_0=0.742-0.748$. The only exception
is the Galactic Bulge, where variables with $P_1/P_0$ as low as
0.726 are found (\cite[Soszy\'nski et al. 2011a]{BulgeRR}). These
low period ratios indicate higher metallicities, up to $Z\sim
0.008$.

{\underline{\it 1O+2O double-mode Cepheids.}} Another type of
double-mode pulsations was detected a few years later in the short
period Cepheid CO\thinspace Aur (\cite[Mantegazza 1983]{COAur}). In
this star the period ratio is very different, $P_2/P_1=0.801$,
implying pulsations in the first (1O) and the second (2O) radial
overtones. Currently, about 500 double-mode Cepheids of this type
are known, most in the Magellanic Clouds (\cite[Soszy\'nski et al.
2008b]{LMCCep}, \cite[2010a]{SMCCep}, \cite[2012]{SEP},
\cite[Marquette et al. 2009]{EROSCep}). Only 19 are identified in
the Galaxy (\cite[Smolec \& Moskalik 2010]{SM10} and references
therein; \cite[Soszy\'nski et al. 2011b]{BulgeCep}). One such
variable has also been found in a spiral galaxy M31 (\cite[Poleski
2013a]{M31Cep}).

\section{New types of multi-mode radial pulsators}

In the next 20 years following discovery of CO\thinspace Aur, no new
types of multi-mode radial pulsators were identified. This started
to change only with the advent of microlensing surveys, which
collected photometric data for thousands of Cepheids and
RR\thinspace Lyrae-type stars. Such data allowed finding rare forms
of oscillations. In just a few years, the inventory of multi-mode
radial pulsators was enlarged by several new classes. Further
discoveries came from analysis of ultra-precise photometry obtained
with {\it Kepler} and CoRoT planet-hunting space telescopes.

{\underline{\it 1O+3O double-mode Cepheids.}} Cepheids pulsating in
the first (1O) and the third (3O) radial overtones are extremely
rare. Only two examples are known, both in the Large Magellanic
Cloud (hereafter LMC, \cite[Soszy\'nski et al. 2008a]{LMCTripl}).
Both variables have period ratio of $P_3/P_1=0.677$. The amplitude
of the third overtone is very low, in both cases below
0.03\thinspace mag in the {\it I}-band.

{\underline{\it F+2O double-mode RR\thinspace Lyrae-type stars.}}
This type of pulsation is also very rare and so far has been
detected only in 9 stars (see \cite[Moskalik 2013]{GranRev} and
references therein). All but one have been discovered with
photometry obtained from space. The ratio of the two pulsation
periods falls in a very narrow range of $P_2/P_0=0.582-0.593$,
implying pulsations in the radial fundamental mode (F) and the
second radial overtone (2O). In most stars of this class the
fundamental mode is modulated, i.e. displays a Blazhko effect.
However, this is not a rule. In two variables with the longest
pulsation periods (V350\thinspace Lyr and KIC\thinspace 7021124) the
amplitudes are constant.

\begin{table}[b]
  \begin{center}
  \caption{Triple-mode Cepheids.}
  \label{tab1}
  {\scriptsize
  \begin{tabular}{lccccccccccl}
  \hline\noalign{\vskip 2pt}
  {\bf ~star}        &~~~& $P_0$\thinspace [day]
                               & $P_1$\thinspace [day]
                                        & $P_2$\thinspace [day]
                                                 & $P_3$\thinspace [day]
                                                          && $P_1/P_0$ & $P_2/P_1$ & $P_3/P_2$ && ref.\\
  \hline\noalign{\vskip 1pt}
  \multicolumn{12}{p{12.5cm}}{\centerline{1O+2O+3O Cepheids}} \\
  \noalign{\vskip -9pt}\hline\noalign{\vskip 1pt}
  ~OGLE-LMC-CEP-1847 &&        & 0.5795 & 0.4666 & 0.3921 &&           & 0.8052    & 0.8404    && 1,2 \\
  ~OGLE-LMC-CEP-2147 &&        & 0.5413 & 0.4360 & 0.3663 &&           & 0.8056    & 0.8401    && 1,2 \\
  ~OGLE-LMC-CEP-3025 &&        & 0.5687 & 0.4582 & 0.3850 &&           & 0.8057    & 0.8402    && 2   \\
  ~OGLE-SMC-CEP-3867 &&        & 0.2688 & 0.2174 & 0.1824 &&           & 0.8086    & 0.8392    && 3   \\
  ~OGLE-BLG-CEP-16   &&        & 0.2955 & 0.2340 & 0.1951 &&           & 0.7919    & 0.8337    && 4   \\
  ~OGLE-BLG-CEP-30   &&        & 0.2304 & 0.1830 & 0.1522 &&           & 0.7943    & 0.8316    && 4   \\
  \noalign{\vskip -1pt}\hline\noalign{\vskip 1pt}
  \multicolumn{12}{p{12.5cm}}{\centerline{F+1O+2O Cepheids}} \\
  \noalign{\vskip -9pt}\hline\noalign{\vskip 1pt}
  ~OGLE-LMC-CEP-0857 && 1.4631 & 1.0591 & 0.8552 &        && 0.7239    & 0.8075    &           && 2   \\
  ~OGLE-LMC-CEP-1378 && 0.5150 & 0.3849 & 0.3094 &        && 0.7474    & 0.8037    &           && 2   \\
  ~OGLE-SMC-CEP-1077 && 0.8973 & 0.6565 & 0.5289 &        && 0.7316    & 0.8056    &           && 3   \\
  ~OGLE-SMC-CEP-1350 && 0.9049 & 0.6660 & 0.5379 &        && 0.7360    & 0.8076    &           && 3   \\
  \noalign{\vskip -1pt}\hline\noalign{\vskip 1pt}
  \multicolumn{12}{p{12.5cm}}{{\it References:} 1 -- \cite[Moskalik et al. (2004)]{MKM04},
                                              2 -- \cite[Soszy\'nski et al. (2008a)]{LMCTripl},
                                              3 -- \cite[Soszy\'nski et al. (2010a)]{SMCCep},
                                              4 -- \cite[Soszy\'nski et al. (2011b)]{BulgeCep}.} \\
  \end{tabular}
  }
  \end{center}
\end{table}

{\underline{\it Triple-mode Cepheids.}} First two variables of this
type were identified by \cite[Moskalik et al. (2004)]{MKM04}.
Currently, only 10 triple-mode Cepheids are known in both Magellanic
Clouds and in the Galaxy (\cite[Soszy\'nski et al. 2008a]{LMCTripl},
\cite[2010a]{SMCCep}, \cite[2011b]{BulgeCep}). They can be divided
into two classes: those which pulsate in the fundamental mode and
the first two radial overtones (F+1O+2O type; 4 objects) and those
which pulsate in the first three radial overtones (1O+2O+3O type; 6
objects). Properties of all known triple-mode Cepheids are
summarized in Table\thinspace\ref{tab1}. These rare objects are very
important and interesting targets for asteroseismology. Modeling of
their three pulsation periods strongly constrains parameters of
these stars and, consequently, imposes tight constraints on Cepheid
evolutionary tracks (\cite[Moskalik \& Dziembowski 2005]{TriplSei}).

{\underline{\it Blazhko effect in 1O+2O double-mode Cepheids.}}
Long-term periodic amplitude modulation was detected in many 1O+2O
double-mode Cepheids of the LMC (\cite[Moskalik et al. 2004]{MKM04},
\cite[2006]{MKM06}, \cite[Moskalik \& Ko{\l}aczkowski 2009]{MK09}).
The behaviour is quite common and occurs in at least 20\% (but
possibly in as many as 35\%) of these stars. Both overtones vary
with the common period ($P_B > 700$\thinspace d), but variations are
anticorrelated -- the maximum amplitude of one mode coincides with
minimum amplitude of the other. The phases (or equivalently periods)
of both modes are modulated, too. The phenomenon closely resembles
the Blazhko effect in the RR\thinspace Lyrae-type star. The only
difference is that not one, but two radial modes are being
modulated. Within accuracy of the data, this modulation is periodic,
and the modulation period is the same for both overtones.

\section{Nonradial modes in Cepheids and RR\thinspace Lyrae-type stars}

Excitation of nonradial modes is very common in pulsating stars.
Modes of this type are observed in all classes of main-sequence
pulsators (\cite[Gautschy \& Saio 1996]{GS96}) as well as in
oscillating white dwarfs (\cite[Winget \& Kepler 2008]{WK08},
Fontaine, these proceedings), subdwarfs (Randall, these proceedings)
and in stochastically driven subgiants and red giants (Hekker, these proceedings).
But classical pulsating variables -- Cepheids and
RR\thinspace Lyrae-type stars -- seemed to be different. They seemed
to avoid excitation of any nonradial oscillations at all. This
picture has changed only in the last decade. New observations prove
that nonradial modes are present also in the classical pulsators.

\subsection{Nonradial modes in LMC Cepheids}

\begin{figure}[b]
 \begin{center}
  \includegraphics[width=13.0cm]{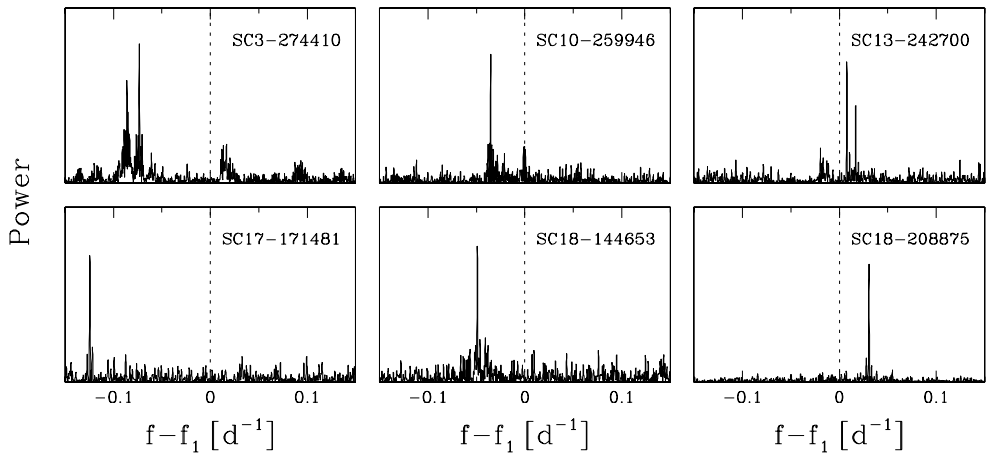}
  \caption{Nonradial modes in LMC first-overtone Cepheids.
           Prewhitened frequency spectra after subtracting the
           dominant radial mode (indicated with dashed line) are
           displayed.}
  \label{fig1}
 \end{center}
\end{figure}

The first convincing detection of nonradial modes in classical Cepheids
came with the analysis of LMC photometry of the OGLE-II survey
(\cite[Moskalik et al. 2004]{MKM04}, \cite[Moskalik \&
Ko\-{\l}acz\-kowski 2008]{MK08}, \cite[2009]{MK09}). In 37 first
overtone Cepheids, low amplitude secondary frequencies have been
found in close proximity to the dominant radial mode ($\Delta f <
0.125$\thinspace d$^{-1}$). Six of these variables are displayed in
Fig.\,\ref{fig1}. Similar secondary frequencies have also been
detected in two F+1O double-mode Cepheids, where they are found
close to the first radial overtone. In most cases only one secondary
peak is present, but sometimes two peaks are seen. In the latter
case, both peaks always appear on the same side of the radial mode.
The observed frequency pattern cannot be explained by any form of
modulation of the radial mode. Modulation should always produce an
equally spaced frequency multiplet centered on the primary peak, and
such a structure is not found in any of these stars. Because
secondary frequencies are too close to the frequency of the radial
first overtone, they cannot be explained by any radial mode, either.
Consequently, they must correspond to nonradial modes of
oscillations.

\begin{figure}[b]
 \vspace*{3.85cm}
 \begin{center}
  \includegraphics[width=8.7cm]{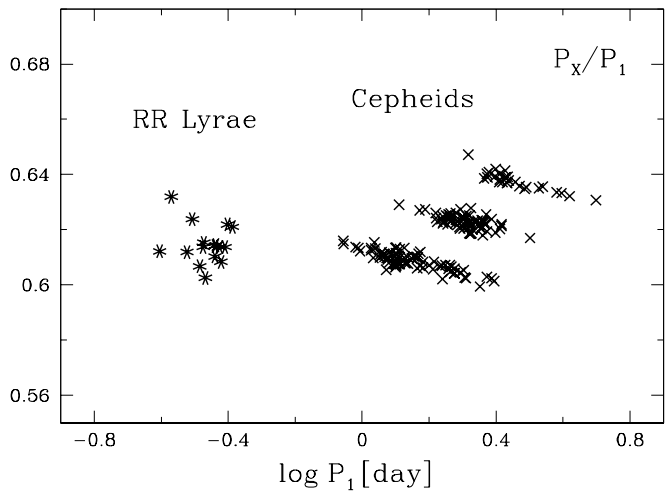}
  \caption{Petersen diagram for Cepheids and RR\thinspace Lyrae-type
           stars with period ratio of $P_X/P_1 = 0.60-0.64$.}
  \label{fig2}
 \end{center}
\end{figure}

\subsection{Double-mode pulsators with period ratio of 0.60\,--\,0.64}

\ \ {\underline{\it Cepheids.}} In some Magellanic Cloud Cepheids a
different type of multiperiodicity is found -- a single secondary
mode is detected with frequency much above that of the radial mode.
The ratio of the secondary and the primary period falls in a narrow
range of $P_X/P_1 = 0.60-0.64$. The first eight Cepheids of this
class have been discovered in the LMC by \cite[Moskalik \&
Ko{\l}aczkowski (2008)]{MK08}. Currently, more than 170 such
pulsators are known in both Magellanic Clouds (\cite[Soszy\'nski et
al. 2008b]{LMCCep}, \cite[2010a]{SMCCep}, \cite[Moskalik \&
Ko{\l}aczkowski 2009]{MK09}). The phenomenon is restricted to the
first overtone Cepheids and F+1O double-mode Cepheids (only one
star). When plotted on the period ratio vs. period diagram (so
called Petersen diagram), these variables form three very tight
parallel sequences (see Fig.\,\ref{fig2}). They are as tight as in
case of the radial double-mode pulsators. However, none of the
sequences fits theoretical prediction for the two radial modes
(\cite[Dziembowski \& Smolec 2009]{DS09}, \cite[Dziembowski
2012]{WD12}). This implies that the secondary mode is nonradial.
Considering that this mode has to be unstable, the theoretical
analysis suggests that it must be an f-mode of high angular degree
of $\ell = 42-50$ (\cite[Dziembowski 2012]{WD12}).

{\underline{\it RR\thinspace Lyrae-type stars.}} Secondary modes
with the period ratio of $0.60-0.64$ are found in RR\thinspace
Lyrae-type stars as well. 16 such variables have been identified so
far, including four stars observed by the {\it Kepler} space
telescope (\cite[Moskalik et al. 2013]{KepRRc0}, \cite[2014]{KepRRc}
and references therein). Most of these variables are dominated by
the first radial overtone (RRc stars), but two are double-mode
pulsators (RRd stars). The amplitude of the secondary mode is always
extremely low, in the mmag range. Interestingly, this mode is found
in all four RRc stars observed by {\it Kepler}. This suggests that
such double-mode pulsations are rather common and should be detected
in many more RRc variables, provided high quality photometry is
available. Just like in the case of Cepheids, comparison of observed
period ratios with theoretical models implies that the secondary
frequency must correspond to a nonradial mode (\cite[Moskalik et al.
2014]{KepRRc}).

The RR\thinspace Lyrae-type stars displaying the puzzling period
ratio of $0.60-0.64$ are very similar to their Cepheid siblings. In
both cases this phenomenon occurs only in the first-overtone
pulsators and in the F+1O double-mode pulsators. The period ratios
fall in the same narrow range. This is shown in Fig.\,\ref{fig2},
where we plot both groups of variables on a common Petersen diagram.
The only difference between the two groups is that classical
Cepheids split into three separate sequences, whereas the
RR\thinspace Lyrae-type stars do not. Finally, in both types of
stars the secondary mode is nonradial. These similarities lead to
the conclusion that Cepheids and RR\thinspace Lyrae-type stars with
the period ratio of $P_X/P_1 = 0.60-0.64$ form a very homogenous
group, constituting a new, well defined class of multimode
pulsators.

\begin{figure}[b]
 \begin{center}
  \includegraphics[width=13.0cm]{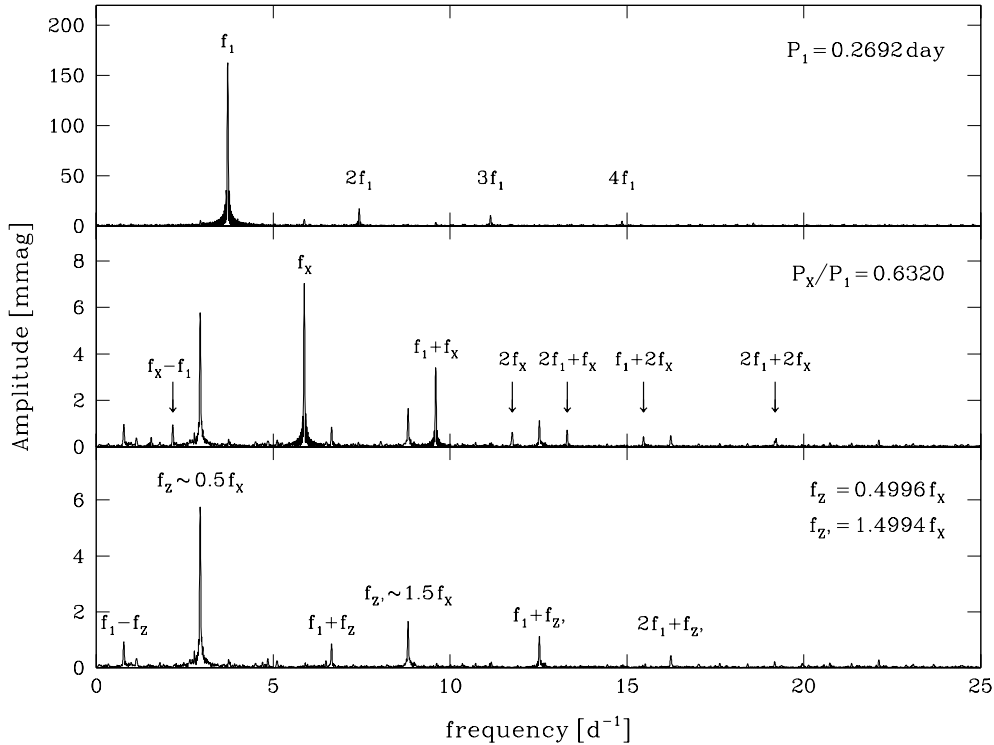}
  \caption{Prewhitening sequence for KIC\thinspace 5520878. Upper
           panel: frequency spectrum of original data. Middle panel:
           frequency spectrum after subtracting $f_1$ and its
           harmonics. Bottom panel: frequency spectrum after
           subtracting $f_1$, $f_X$ and their harmonics and linear
           combinations.}
 \label{fig3}
 \end{center}
\end{figure}

{\underline{\it Period doubling of the secondary mode in
RR\thinspace Lyrae-type stars.}} In Fig.\,\ref{fig3} we display a
pre\-whitening sequence for KIC\thinspace 5520878, one of the RRc
stars in the {\it Kepler} field (\cite[Moskalik et al.
2014]{KepRRc}). After subtracting the dominant radial pulsation
($f_1$) and its harmonics, the frequency spectrum of the residuals
(middle panel) reveals a secondary frequency at $f_X =
5.879$\thinspace d$^{-1}$, which yields a period ratio of $P_X/P_1 =
0.632$. A harmonic of $f_X$ and several linear combinations with
$f_1$ are also present. But ultra-precise {\it Kepler} photometry
reveals even more. After prewhitening the data with $f_1$, $f_X$ and
their harmonic and combination frequencies (bottom panel), the
highest residual peak appears at $f_Z = 2.937$\thinspace d$^{-1}$,
that is at $\sim\! 1/2 f_X$. In other words, $f_Z$ is not an
independent frequency, but a subharmonic of $f_X$. Another
subharmonic is found at $f_{Z'}\sim 3/2 f_X$. Detection of
subharmonic frequencies is highly significant. Their presence is a
characteristic signature of a period doubling of the secondary mode,
$f_X$. The same signature of the period doubling behaviour is seen
in all four RRc variables in the {\it Kepler} field (\cite[Moskalik
et al. 2013]{KepRRc0}, \cite[2014]{KepRRc}). It is also detected in
an RRd variable AQ\thinspace Leo (\cite[Gruberbauer et al.
2007]{AQLeo}). Thus, RR\thinspace Lyrae-type stars with the period
ratio of 0.60\,--\,0.64 are yet another class of pulsators in which
period doubling can occur. We recall, that this phenomenon has
recently been discovered in two other classes of pulsating variables
-- in Blazhko RRab stars (\cite[Szab\'o et al. 2010]{PDRRab}) and in
the BL\thinspace Herculis-type stars (\cite[Smolec et al.
2012]{BLHerPD}). Its origin can be traced to a half-integer
resonance between the pulsation modes (\cite[Moskalik \& Buchler
1990]{MB90}).

\subsection{Other nonradial modes in RRc stars}

With the benefit of high quality {\it Kepler} data it is possible to
search for modes which are even weaker than $f_X$ and its
subharmonic peaks. Such modes are discovered in three {\it Kepler}
RRc variables (\cite[Moskalik et al. 2014]{KepRRc}). All but one of
these oscillations have amplitudes below 0.4\thinspace mmag.
The majority of them generate combination peaks with the dominant radial
pulsation ($f_1$), which proves that they indeed originate in the
RRc star and do not come from blending with another variable.

\begin{figure}[b]
 \begin{center}
  \includegraphics[width=13.0cm]{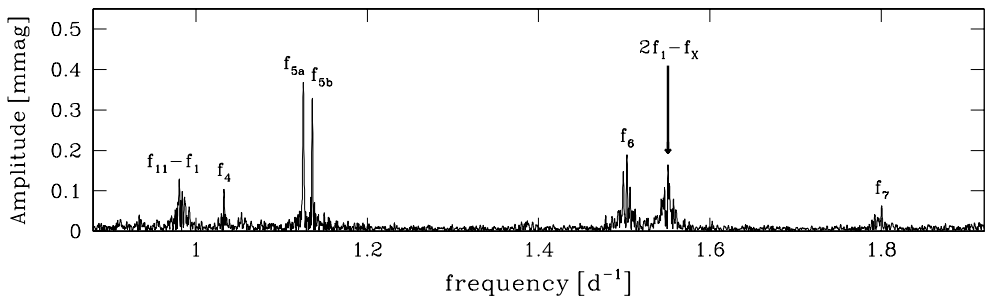}
  \caption{Low frequency low amplitude modes in KIC\thinspace
           5520878. Fourier transform of Q1 -- Q10 lightcurve is
           computed after prewhitening the data of the dominant
           frequency $f_1$, $f_X$, their harmonics, subharmonics and
           5 highest combination peaks. Nonradial modes appear at
           $f_4$, $f_5$ (doublet), $f_6$ (triplet) and $f_7$.
           $f_{11}\! -\! f_1$ is a combination peak associated with
           another low amplitude mode. The frequency of the radial
           fundamental mode is estimated to be $f_0 \sim
           2.75$\thinspace d$^{-1}$.}
  \label{fig4}
 \end{center}
\end{figure}

The richest harvest of low amplitude modes is found in KIC\thinspace
5520878, where 15 such oscillations are detected. Based on the
period ratios, one of these modes can be identified with the second
radial overtone, but all others must be nonradial. Interestingly,
several modes have frequencies significantly lower than the radial
fundamental mode (see Fig.\,\ref{fig4}). This implies that these
nonradial oscillations are not $p$-modes, but must be of either
gravity or mixed mode character. Low frequency modes of this type
are discovered also in the other two RRc stars.

\subsection{Nonradial modes in RRab stars}

Low amplitude modes other than the radial second overtone are
detected in six RRab variables. They are listed in
Table\thinspace\ref{tab3}. For each secondary mode, instead of its
period, we provide the period ratio of this mode and the dominant
fundamental radial pulsation, $P/P_0$. The period ratios given in
the column 4 of the Table are clearly incompatible with those of the
radial modes. Therefore, these secondary oscillations must be
nonradial. The situation is somewhat more complicated for modes
listed in column 3. Their period ratios are more or less similar to
that observed in the RRd stars. On this ground, the secondary modes
of column 3 are usually identified with the radial first overtone.
However, the double-mode RR\thinspace Lyrae-type stars do not
populate the $P/P_0$ vs. $P_0$ diagram in a random way. Instead,
they form a rather narrow, well defined sequence on this diagram
(\cite[Soszy\'nski et al. 2009]{LMCRR}, \cite[2011a]{BulgeRR}).
Putting this differently, there is an empirical relation between the
period ratio and the period. It is easy to check that
V1127\thinspace Aql is the only variable of
Table\thinspace\ref{tab3} which follows this relation. For all the
remaining variables, the secondary modes listed in column 3 deviate
considerably from the empirical trend. Therefore, identification of
these modes with the radial first overtone should be treated with
great caution. These periodicities most likely correspond to
nonradial modes, the frequencies of which are just close to that of
the radial overtone. We note, that a small number of similar
outliers from the empirical $P/P_0$ vs. $P_0$ relation have also
been identified in the LMC and in the Galactic bulge
(\cite[Soszy\'nski et al. 2009]{LMCRR}, \cite[2011a]{BulgeRR}). The
secondary periods in those objects might correspond to nonradial
modes, too.

The tendency of nonradial oscillations to appear preferentially
around the frequency of the radial overtone is not a surprise. In
fact, it is consistent with the theory. Linear nonadiabatic
calculations show that nonradial modes can be excited in the
RR\thinspace Lyrae-type stars, and their growth rates are highest in
the vicinity of the radial modes (\cite[Van Hoolst et al.
1998]{VHDK98}, \cite[Dziembowski \& Cassisi 1999]{DC99}). The growth
rates reach maximum values at frequencies slightly above that of the
radial mode. This is true both for modes of $\ell =1, 2$ and for the
strongly trapped modes (STU modes) of higher spherical degrees.
Thus, nonradial oscillations which are most likely to be excited
should yield period ratios somewhat lower than the radial first
overtone. This prediction is in good agreement with observational
picture summarized in Table\thinspace\ref{tab3}.

\begin{table}[t]
  \begin{center}
  \caption{Secondary periodicities in RRab stars}
  \label{tab3}
  {\scriptsize
  \begin{tabular}{lcccl}
  \hline\noalign{\vskip 2pt}
  {\bf ~star}                  & $P_0$\thinspace [day]
                                        & $P/P_0$ & $P/P_0$
                                                            & ~ref.\\
  \noalign{\vskip 0pt}\hline\noalign{\vskip 1pt}
  ~V1127\thinspace Aql         & 0.3560 & 0.7271  & 0.6966  & ~1,2 \\
  ~RR\thinspace Lyr            & 0.5669 & 0.7557  &         & ~5   \\
  ~V354\thinspace Lyr          & 0.5617 & 0.7295  & 0.8555  & ~3   \\
  ~V360\thinspace Lyr          & 0.5576 & 0.7210  &         & ~3   \\
  ~V445\thinspace Lyr          & 0.5131 & 0.7306  & 0.7031  & ~4   \\
  ~CoRoT\thinspace 105288363   & 0.5674 & 0.7216  & 0.7764  & ~4   \\
                               &        & 0.7407  &         &      \\
  \noalign{\vskip -1pt}\hline\noalign{\vskip 1pt}
  \end{tabular}
  }
  \end{center}
 \vspace{1mm}
 \scriptsize{
 {\it References:} 1 -- \cite[Chadid et al. (2010)]{V1127Aql},
                   2 -- \cite[Poretti et al. (2010)]{Poretti},
                   3 -- \cite[Benk\H{o} et al. (2010)]{Benko},
                   4 -- \cite[Guggenberger et al. (2012)]{V445Lyr},
                   5 -- \cite[Moln\'ar et al. (2012)]{RRLyr}.
 }
\end{table}

\begin{acknowledgements}
This research is supported by the Polish National Science Centre
through grant DEC-2012/05/B/ST9/03932. Generous IAU support in the
form of a travel grant is also acknowledged.
\end{acknowledgements}

\end{document}